# Evolution and defect analysis of vertical graphene nanosheets

Subrata Ghosh[1,*], K. Ganesan[1,†], Shyamal R. Polaki[1], T.R. Ravindran[1], Nanda Gopala Krishna[2], M. Kamruddin[1] and A.K. Tyagi[1]

[1] Materials Science Group, Indira Gandhi Centre for Atomic Research, Kalpakkam - 603102, India.

[2] Corrosion Science and Technology Group, Indira Gandhi Centre for Atomic Research, Kalpakkam - 603102, India.

## Abstract

We report catalyst-free direct synthesis of vertical graphene nanosheets (VGNs) on $SiO_2$/Si and quartz substrates using microwave electron cyclotron resonance - plasma enhanced chemical vapor deposition. The evolution of VGNs is studied systematically at different growth stages. Raman analysis as a function of growth time reveals that two different disorder-induced competing mechanisms contributing to the defect band intensity. The VGNs grown on $SiO_2$/Si substrates predominantly consists of both vacancy-like and hopping defects. On the other hand, the VGNs grown on quartz substrates contain mainly boundary-like defects. XPS studies also corroborate Raman analysis in terms of defect density and vacancy-like defects for the VGNs grown on $SiO_2$/Si substrates. Moreover, the grown VGNs exhibit a high optical transmittance from 95 to 78 % at 550 nm and the sheet resistance varies from 30 to 2.17 k$\Omega$/□ depending on growth time.

## 1. Introduction

Over the past few decades, carbon nanostructures opened a new window to scientific community due to their unique structures and exotic physical properties.[1-3] The graphene, newborn in carbon wonderland, has fuelled academic and industrial interest due to its unusual electronic and mechanical properties, in particular, ballistic and tunable transport properties.[3-5]

---

[*] Electronic mail : subrataghosh.phys@gmail.com (Subrata Ghosh)
[†] Corresponding author. Tele.: +91-44-27480500 Extn: 22514. Email : kganesan@igcar.gov.in ( K. Ganesan)

To realize its potential for future nanoelectronics, a reliable, fast and reproducible synthesis technique is necessary. The mass production of large area graphene is the most challenging goal for device applications. The general approaches to graphene synthesis are exfoliation (mechanical or chemical), epitaxial growth, graphene oxide (GO) reduction and chemical vapor deposition (CVD).[6] Among them, thermal CVD is the most popular technique to overcome the limitations of scalability. The interaction between carbon π orbital and surface atoms of the metal substrates (Cu, Ni) plays a major role in graphene growth.[5,7] However, the CVD grown graphene has to undergo a major hurdle of transferring onto a dielectric substrate before entering into devices. The transfer process inevitably introduces impurities and it could degrade the quality of the graphene. For example, the organic solvent polymethyl methacrylate (PMMA) used in transfer can be a source of *p*-type doping element, and also dissolving the PMMA residue completely from graphene is not easy.[8]

Catalysis-free direct synthesis of graphene on dielectric substrates, compatible with complementary metal oxide semiconductor technology, is a challenging task. Recently, the direct growth of graphene on dielectric substrates are explored by thermal CVD, yet with limited success.[9,10] Alternatively, the plasma enhanced CVD (PECVD) have already shown its potential in growing various carbon nanostructures. The main advantages of PECVD are that it is catalyst free, short growth time can be carried out at relatively low temperatures and is amenable to doping. In addition, the technique is ideal for uniform large area and conformal coatings. There are several reports on growth of carbon nanostructures by various plasma based techniques such as microwave PECVD, dc plasma discharge, ICP-plasma and thermal plasma jet systems.[11-20] However, the PECVD grown samples are found to be vertically oriented which are attributed to the in-built electric fields associated with plasma.[15] Due to the structural characteristics with high aspect ratio and large 3D networks, the vertical graphene nanosheets (VGNs), also called as carbon/graphene nanowalls/nanosheets/naoflakes, are excellent candidates for applications in field emission, fuel cells , chemical & biosensors and energy storage devices.[12,14,16,17] In order to use them in an efficient way, a detailed understanding of growth mechanism and their structural evolution are necessary.

Earlier studies on growth of carbon nanotubes by plasma based techniques using metallic catalyst lead to the invention of VGNs.[15] Several groups have explored the growth of graphene/carbon nanosheets on catalytic as well as non-catalytic metal surfaces. However, synthesis of VGNs on semiconducting and dielectric substrates is rare. In an excellent review, Bo et al[11] summarized the influence of various key process parameters such as type of feedstock gas and composition, microwave power, electric field, temperature and pressure on the synthesis of VGNs and also the growth model for the vertical orientation of nanosheets on various types of substrates. However, it should be noted that still, there is no unified theory to unveil the growth mechanism and to optimize the process parameters for a given plasma source.[11] Further, the PECVD synthesized materials suffer from large amount of defects due to ion bombardment from the high density plasma. Hence, it is very important to gain knowledge over the defect formation mechanism with respect to the process parameters in order to minimize defects. Raman spectroscopy is an excellent tool to investigate the defects in carbon materials. Though almost all previous reports on plasma based graphene synthesis include Raman data, a detailed analysis of defects and disorder, such as nature of defects, quantification of defect concentration and their correlation with growth mechanism, are scarce.

In view of proven capability to synthesis electronic grade carbon materials, we consider microwave electron cyclotron resonance plasma enhanced chemical vapor deposition (ECR-CVD) as a promising technique to fabricate VGNs on $SiO_2$/Si and quartz substrates.[17] In the present study, we discuss the catalyst-free synthesis of VGNs and their structural analysis using scanning electron microscopy (SEM), Raman spectroscopy and x-ray photoemission spectroscopy (XPS) techniques. A quantitative defect analysis is carried at various stages of growth and the types of defects associated with the two different substrates are identified. XPS studies are also carried out to reveal surface chemistry and support defect analysis. Further, electrical and optical properties of VGNs are studied at room temperature. Overall, a comprehensive understanding on the growth mechanism of VGNs on $SiO_2$/Si and quartz substrates and its defects analysis are presented.

## 2. Experimental methods

### 2.1 Growth of VGNs by ECR-CVD

The VGNs are deposited on $SiO_2$/Si and quartz substrates under $CH_4$/Ar plasma in an ECR-CVD reactor. Ar is chosen as a dilution gas since it enhances the $C_2$ radical production which helps to grow high quality graphitic VGNs. Also, Ar+ ions in plasma enhance the electron energy and hence improve the chemical reactivity and plasma stability.[11] The schematic of the ECR-CVD system is given in Fig. 1. The system consists of a microwave source operating at 2.45 GHz with maximum power of 1.2 KW and an electromagnet generating a magnetic field of 875 Gauss which satisfies the ECR condition. The source gas $CH_4$ is sent through a circular showerhead that uniformly injects gas into the reaction chamber. An adjustable substrate holder with heater allows manipulation of the distance between the quartz window and the substrate. Growth time ranges from 10 sec to 45 min. Ultrasonically cleaned substrates are loaded into the substrate holder and the chamber is evacuated down to a base pressure $5 \times 10^{-6}$ mbar by a turbomolecular pump. The substrate temperature is increased to 750 °C and after 20 minutes, the substrates are pre-cleaned by Ar plasma under 20 sccm flow rate for 10 min at 200W microwave power. After pre-cleaning the substrates, $CH_4$ is fed into the chamber through circular showerhead at 2.5 sccm flow rate, subsequently Ar flow is reduced to 2.5 sccm. The growth experiments are then performed at the microwave power of 400 W. The operating pressure of the reactor chamber is maintained about $1 \times 10^{-3}$ mbar. Following the growth, the plasma is turned off and the substrates are annealed for 10 min at the same growth temperature. Finally, the samples are cooled down to room temperature and taken out for characterization.

### 2.2 Characterization

The VGNs morphology is studied using a field emission scanning electron microscope (SEM, Supra 55, Zeiss). The structural properties, in terms of defects and disorder, are evaluated by micro-Raman spectroscopy (Renishaw inVia, UK) with 514 nm laser and accumulation time of 10 seconds with 100X objective lens. In order to avoid the laser induced heat on the sample, the laser power was kept below 1 mW. Surface chemical analysis of the VGNs are carried out by x-ray photoelectron spectroscopy (XPS, M/s SPECS, Germany). Sheet resistance is measured in van der Paw geometry using Agilent B2902A precision source/measure unit. The optical transmittance is measured by ultraviolet-visible spectroscopy (UV-Vis, Avantes AvaLight-DH-S-BAL) in the range 200-800 nm for the VGNs grown on quartz substrates.

## 3. Results and discussion
### 3.1 SEM analysis

The surface morphologies of the VGNs grown on $SiO_2$/Si substrates are characterized by SEM at different stages of growth (Fig. 2 (a-d)). Fig. 2e shows the cross sectional SEM image of VGNs grown for 45 minutes and it has the height of 48 nm. The variation in number density and height with duration of growth is given in Fig 2f. At the very early stages of growth, of about 10 seconds duration, the film consists of many island like structures over the smooth substrate as seen in Fig. 2a. It should be noted that even at this very short duration of growth, the film is electrically conductive with sheet resistance of about 30 k$\Omega$/□. This indicates the initial formation of nanographite islands and they coalesce to form electrically continuous nanographite layer parallel to the substrate surface. In the meantime, as an outcome of coalescence, stress is relieved at the interface of the nanographite grain boundaries which results in the nucleation of carbon in the vertical direction.[19] Further increase in growth time, brings more carbon ad-atoms to the growing film and carbon diffusion takes place in the planar regions as well as the vertically growing graphene nanosheets. However, the carbon atoms diffusing to the edge of vertical graphene binds well due to strong in-plane C-C covalent bond and grow normal to the substrate. On the other hand, the carbon species arriving at planer region are re-evaporated due to the weak van der Waals force between the graphene layers. Meanwhile, the atomic hydrogen generated from $CH_4$ by microwave energy also etches out amorphous carbon (a-C), $sp^3$-C and $sp^2$-C at different rates. This simultaneous growth and etching process makes the VGNs to increase the height and decrease the number density with growth time as seen in Fig 2 (a-f). Further, a detailed Raman and XPS analysis are carried out to substantiate the growth mechanism as discussed below.

## 3.2 Raman analysis

Raman spectroscopy is a well-established and powerful technique for structural characterization of carbon-based materials. In addition to ascertaining the crystalline state, it is useful to investigate defects and disorder in the family of carbon materials.[13,21,23-30] Here, we analyze the abundance and nature of defects formed during the evolution of VGNs on $SiO_2$/Si and quartz substrates. Fig. 3 shows a typical Raman spectrum of VGNs grown by ECR-CVD. The spectrum consists of two prominent bands - G and G′ (also called as 2D ) at ~ 1580 and 2700 cm$^{-1}$ respectively, which are characteristic to graphitic structure. Also, it consists of one phonon defect-assisted processes such as D, D′ and D″ bands and two phonon defect-assisted processes such as 2D′, D+D′ and D+D″ bands. The D, D′, 2D′, D″, D+D′ and D+D″ bands are activated only by defects and these bands are absent in pristine graphene. Also, these bands are strongly dispersive with excitation energy due to the presence of defects and disorder.[21] The D band (~1350 cm$^{-1}$) and D′ band (~ 1620 cm$^{-1}$) respectively originate from the process of inter-valley and intra-valley double resonance (DR). The 2D′ band is the overtone of D′ band and appears at ~ 3250 cm$^{-1}$. The peak at ~ 2485 cm$^{-1}$ is assigned to D+D″ band which is due to the combination of D phonon and the zone boundary phonon corresponding to LA branch at 1100 cm$^{-1}$.[21] The weak and broad band around 1100 cm$^{-1}$, as shown in inset of Fig. 3a, could arise from hydrogenated carbon traces associated with a bond stretching mode of sp$^3$ sites.[22]

Fig. 4a shows the Raman spectra of VGNs grown on $SiO_2$/Si substrates with different growth time ranging from 10 sec to 45 minutes. Similar spectra observed for films grown on quartz substrates are not shown here. The Raman spectrum of the film grown for 10 seconds shows a very broad D and G bands with nearly negligible G′ band intensity indicative of a more disordered structure. The intensity of the D, G and G′ bands increases with growth time. Also, the FWHM of D, G and G′ bands decreases with growth time (shown in Fig. 4b). A significant intensity in G′ band is observed when the growth time exceeds 10 minutes as shown in Fig. 4a. Further, the increase in growth time shifts the G band position [Pos(G)] from 1598.0 (1598.7) to 1590.7 (1588.8) cm$^{-1}$ for the VGNs grown on $SiO_2$ (quartz) substrates indicating the improvement in crystallinity (Fig. 4c).[23] The other two bands, D and G′, do not show any systematic change in peak position with respect to growth time. Also, the intensity ratio of G′ to G band ($I_{G'}/I_G$) increase with respect to growth time as shown in Fig 4c. The observed FWHM of G and G′ bands of the films grown for 45 mins on $SiO_2$/Si (quartz) substrates are about 58 (38) and 95 (80) cm$^{-1}$ respectively. The observed large D band intensity is due to the presence of disorder which arises from large amount of edge states, sp$^3$ bonded C-H species, nanographitic base layer and ion induced defects from the plasma during growth.

Defects play a crucial role in determining the physical and mechanical properties of the mesoscopic structures. In order to engineer the desired properties of graphene, it is important to study the correlation between the type of defects and its Raman scattering process. In recent times, considerable efforts are devoted to investigate the nature of defects and its quantification on graphene by introducing specific type of defects through Ar$^+$-ion bombardment or plasma surface modification.[13,24-28] In a disordered graphitic system, it is known that the intensity ratios of D to G ($I_D/I_G$) provide a measure of defects / disorder in the films. When defect density is relatively large as in the case of present study, it is more appropriate to take the peak intensity ratio rather than peak area ratio. The latter is more accurate for a system with low defect concentration.[27, 28] According to the resonant Raman scattering theory, the intensity of defect bands ($I_D$, $I_{D'}$) not only depends on the amount of defects but also on the type of defects associated with it. The ratio of $I_D/I_{D'}$ would provide information about the type of defects present in the material. According to classification, if the ratio of $I_D/I_{D'}$ is 13, it indicates the presence of sp$^3$ related defects, and similarly, 10.5 corresponds to hopping defects, 7 for vacancy-like defects, 3.5 for boundary-like defects and 1.3 represents the on-site defects in graphene.[28,29] In fig. 5, the plot of $I_D/I_G$ versus $I_{D'}/I_G$ shows a linear behavior and agree well with the earlier reports.[29] The observed slopes ($I_D/I_{D'}$ ratio) are found to be 8.6±2.9 and 4.0 ± 0.7 for the VGNs grown on $SiO_2$/Si and quartz substrates respectively. The slopes indicate the existence of combined vacancy-like and hopping defects for the VGNs grown on $SiO_2$/Si substrates and boundary-like defects for the VGNs grown on quartz substrates. The different values of $I_D/I_{D'}$, $I_{G'}/I_G$ and FWHM of phonon modes (D, G, G′) for the VGNs grown on $SiO_2$/Si and quartz are due to dissimilar adsorption abilities of incoming carbon ions on different substrates.[16]

Fig. 6a shows the variation of $I_D/I_G$ with respect to growth time. It indicates that the $I_D/I_G$ ratio has a non-monotonic dependence on growth time, and extends with growth time upto 10 minutes, and then decreases beyond 10 minutes. However, as discussed earlier (Fig 4b), the decreased FWHM of D, G, G′ bands signify the reduction of defects and disorder in the VGNs with growth time. Hence, the non-monotonic behavior of $I_D/I_G$ ratio with growth time suggests the existence of two disorder-induced competing mechanisms contributing to the D band intensity. This behavior can be understood in terms of the well established defects and disorder model, amorphization trajectory of graphitic materials, proposed by Ferrari and Robertson.[30] The transformation of highly $sp^2$ bonded graphite into highly $sp^3$ bonded tetrahedral amorphous carbon (ta-C) occurs through three stages viz. (1) graphite to nanocrystalline graphite [low defect (LD)]; (2) nanocrystalline graphite to low $sp^3$ a-C [high defect (HD)] and (3) low $sp^3$ a-C to high $sp^3$ ta-C. The mechanism for the formation of D band intensity is different in each stage. It is established that the $I_D$ is directly proportional to amount of defects in stage 1 and inversely proportional to defects in stage 2, and also $I_G$ is always proportional to the amount of $sp^2$ rings present in the sample.[25,29] Hence, we attribute the non-monotonic behavior of $I_D/I_G$ with growth time to the transformation of defects and disorder from stage 2 to stage 1. Thus, the VGNs grown for ≤ 10 min and > than 10 minutes follow the stage 2 and stage 1 respectively.

Cancado *et al* [25] developed a model to estimate the inter defect distance ($L_D$) and defect density ($n_D$) in ion bombarded graphene by Raman spectroscopy with appropriate boundary conditions. According to the model, the $L_D$ and $n_D$ can be quantified in stage 1 and stage 2 as given below.

$$L_D^2(LD) = \frac{4.3 \times 10^3}{E_L^4} \left(\frac{I_D}{I_G}\right)^{-1} \quad \ldots\ldots\ldots 1$$

$$L_D^2(HD) = 5.4 \times 10^{-2} E_L^4 \left(\frac{I_D}{I_G}\right) \quad \ldots\ldots\ldots 2$$

$$n_D(LD) = \left(7.3 \times 10^9\right) E_L^4 \left(\frac{I_D}{I_G}\right) \quad \ldots\ldots\ldots 3$$

$$n_D(HD) = \frac{5.9 \times 10^{14}}{E_L^4} \left(\frac{I_D}{I_G}\right)^{-1} \quad \ldots\ldots\ldots 4$$

where $E_L$ is the excitation energy of the laser. The equations 1 and 3 are valid for low (stage 1) defect concentration and the equations 2 and 4 are valid for high (stage 2) defect concentration regimes in the graphitic structure. By considering similar assumption,[25] we calculated $L_D$ ($n_D$) using equations 1(3) and 2(4) for films grown for > 10 mins and ≤ 10 mins respectively. Fig. 6b shows the variation of $I_D/I_G$ with respect to $L_D$, exhibiting a non-monotonic behavior. The mechanism for such behavior is discussed earlier (Fig 6a). The non-monotonic trend was also observed for ion bombarded graphene and explained with a phenomenological model for quantifying ion-induced defects in graphene.[26] According to the model,

$$\frac{I_D}{I_G} = C_A \frac{r_A^2 - r_S^2}{r_A^2 - 2r_S^2} \left[ \exp\left(\frac{-\pi r_S^2}{L_D^2}\right) - \exp\left(-\pi \frac{r_A^2 - r_S^2}{L_D^2}\right) + C_S \left(1 - \exp\left(\frac{-\pi r_S^2}{L_D^2}\right)\right) \right] \quad \ldots\ldots\ldots 5$$

where $C_A$ and $C_S$ are the maximum possible value of $I_D/I_G$ when the hexagonal network of carbon atoms are not disturbed and highly disordered limit respectively. The parameters $r_A$ and $r_S$ are the radius of defect activated region and structurally disordered region respectively (with $r_A > r_S$). The solid line in the Fig. 6b is the fit to the experimental data along with simulated curve with the same fitting parameters ( $0 < L_D < 50$ nm ) and finds an excellent agreement with the model. The best fit parameters are found to be $r_A$=3.81 nm, $r_S$=1.08 nm, $C_A$=5.0 and $C_S$=0.4 which are closely matching to the reported values.[26] The fitted curve indicates an initial increase of $I_D/I_G$ ratio up to a critical $L_D$ value of about 3.8 nm and further monotonic decrease with increase in $L_D$.[25-27] Hence, this plot clearly indicates the transformation of defects and disorder with growth time which demonstrate the cross over from stage 2 to stage 1 of Ferrari and Robertson model.[30] Fig. 6c shows the variation of defect concentration with respect to growth time. As it can be seen from the figure 6c, the defect concentration continuously decreases with growth time. Thus, Raman studies verify the evolution of graphene nanosheets and reveal the type of defects present in two different substrates. It is found that the defect density is so high at early stage of growth due to the nano-graphite island formation. Later, the defect density monotonically decreases due to the formation of high pure VGNs. These observations are in excellent agreement with the growth mechanism as discussed in SEM analysis section 3.1.

### 3.3 XPS analysis

In order to strengthen the growth mechanism obtained from SEM and defect transformation at different stages obtained from Raman spectroscopy, XPS studies are carried out for the VGNs films grown on $SiO_2/Si$ substrates. Fig. 7a shows the high resolution C1s XPS spectra of the films grown for 10 sec, 10 and 45 mins. The XPS spectra show a clear asymmetry in the higher binding energy (B.E.) regime for all the three samples, which is typically observed for high conductive metals and also the asymmetry increases with growth time. These observations confirm that the VGNs are highly conducting and of graphitic nature.[31,32] As seen in the Fig. 7a, the peak position of the C1s spectra are found to shift towards lower B.E. with growth time. The deconvoluted high resolution C1s spectrum of the film grown on $SiO_2/Si$ for 10 mins is shown in Fig. 7b. The main peak ~ 284.4 eV corresponds to C=C $sp^2$ bonded graphitic structure, and the one around 285.1 eV corresponds to defect structure arising from C-H related $sp^3$ bonded structure.[14,17,33] In addition, the peaks related to C-O, C=O bonding, probably from adsorbed impurities are also observed. This C-H related $sp^3$ bonded structure is further confirmed by FTIR spectroscopy which shows absorption peaks at around 2958 and 2875 $cm^{-1}$ corresponding to the asymmetric and symmetric $sp^3$-$CH_3$ vibration respectively (results are not shown here).[34] Herein, there is also one more peak near 284.1 eV which is closely matching to the B.E. for Si-C alloy. In order to verify Si-C alloy formation at $SiO_2$-nanographite interface during the nucleation process, Si2p XPS spectra are taken for all the three samples and are shown in inset of Fig. 7c. As can be seen from Fig 7c, there is no evidence for the formation of Si-C, which should show a signature at around 100.8 eV. It should be noted that Raman measurements also did not show any evidence for Si-C formation, which should have a phonon mode at about 800 $cm^{-1}$. Hence, we exclude the phenomenon of Si-C formation which is reported in some earlier studies.[17, 35]

To account for the peak with chemical shift lower to B.E. of C1s, in the case of muliwalled carbon nanotubes, Barinov *et al* [36] had demonstrated that the peak is due to the vacancies and also they verified it on the intentionally introduced vacancy-like defects in highly oriented pyrolytic graphite. Hence, we assign the peak at around 284.1 eV to the presence of vacancy-like defects. The Raman analysis also indicates the existence of combined vacancy-like defects and hopping defects which is in an excellent agreement with XPS analysis. Moreover, the hopping defects and vacancies are closely related in disordered system because the hopping of charge carriers / atoms can take place on the vacancy sites.[36] As shown in Table 1, the B.E. of all peak positions (C=C $sp^2$, C-H $sp^3$ related defects, C-O, C=O, and vacancy-like defects) in the spectrum decreases with increase in growth time. Also, the area under the curves for defect related peaks continuously decreases which indicates the lowering of defect density with growth time. Further, as seen in inset of Fig. 7a, π-π* energy loss peak increases with growth time indicating the high conductive graphitic structure of the samples with time. These results confirm the evolution of refined graphitic structure with time. Further, electrical and optical studies are carried out and discussed below to support the study.

### 3.4 Electrical and optical properties of VGNs

Fig. 8a shows the room temperature sheet resistance of the VGNs grown on $SiO_2$/Si with respect to growth time. The sheet resistance decreases from 30 to 2.17 kΩ/□ with increasing growth time and it shows an ohmic behaviour in I-V characteristics under low bias condition (as shown in inset of Fig 8a). The decreasing resistance is mainly due to the growth of the high pure and high conductive few layer VGNs networks over the relatively high resistive nanographite film. As the growth time increases, the length and height of the VGNs increases with decrease in defect density as evidenced from Raman and XPS analysis. Hence, the sheet resistance decreases with growth time due to the domination of conducting pure VGNs. The optical transmittance in UV-Vis range from 200 to 800 nm is recorded for the samples grown on quartz substrates and shown in Fig. 8b. The inset in Fig 8b shows the optical transmittance at 550nm over growth time and it varies from 95 to 78 %. The samples grown for 10 sec exhibits very high optical transmittance and it decreases with time due to the increase in thickness as well as the high conducting nature of the VGNs. All the films show a strong absorption at about 270 nm which is due to the π-π* transition in graphitic system and the increased absorption with time indicate improved graphitic nature of the films.

## 4. Conclusion

A catalyst-free direct synthesis of VGNs has been demonstrated by ECR-CVD at a relatively low substrate temperature on $SiO_2$/Si and quartz substrates. Our experimental observations suggest that the growth mechanism is based on the direct adsorption and surface diffusion of carbon species rather than the carbide formation at the interface. For the first time, we have identified the type of defects and the transformation of defects for the VGNs grown by plasma based systems. The transformation of defects with growth time is in conceptual agreement with a well established amorphization trajectory for graphitic materials. The VGNs grown on $SiO_2$/Si substrates contains predominantly combined vacancy-like and hopping defects. On the other hand, the VGNs grown on quartz substrates contain mainly boundary-like defects. Further, the amount of defects considerably decreases with growth time grown on both the substrates. The XPS studies carried out on $SiO_2$/Si substrates also support the existence of vacancy-like defects. Moreover, the area under the curve for XPS defect peaks monotonically decrease with growth time indicates that the defect density decreases with growth time. Further, the grown films show an excellent transmittance and low sheet resistance. The clarity on the growth mechanism, defect analysis and their correlation with electrical and optical properties would help in develop films with desired characteristics which are in demand for optoelectronic and energy applications.

Table1. The parameters extracted from XPS analysis for the VGNs grown on $SiO_2/Si$ substrate

| Growth time | C=C $sp^2$ | | C-H $sp^3$ defect | | C-O | | C=O | | Vacancy-like defects | |
|---|---|---|---|---|---|---|---|---|---|---|
| | BE (eV) | Area (%) | BE (eV) | Area (%) | BE (eV) | Area (%) | BE (eV) | Area (%) | BE (eV) | Area (%) |
| 10 sec | 284.71 | 45.24 | 285.21 | 36.19 | 286.00 | 6.67 | 286.65 | 3.81 | 284.2 | 6.67 |
| 10 min | 284.65 | 59.81 | 285.15 | 24.30 | 285.90 | 6.54 | 286.55 | 3.11 | 284.1 | 4.98 |
| 45 min | 284.55 | 60.12 | 285.04 | 25.30 | 285.75 | 6.54 | 286.45 | 3.12 | 283.9 | 3.57 |

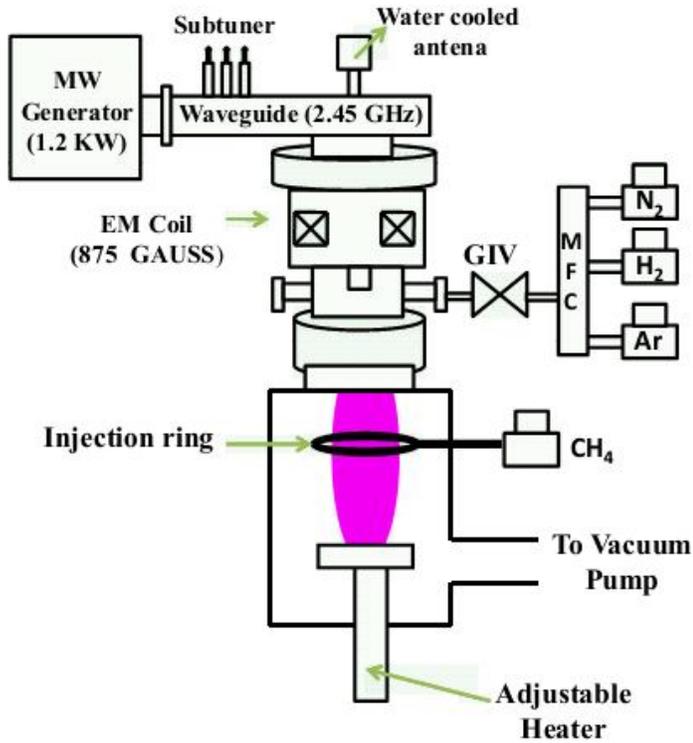

Fig.1- Schematic of the ECR-CVD reactor chamber

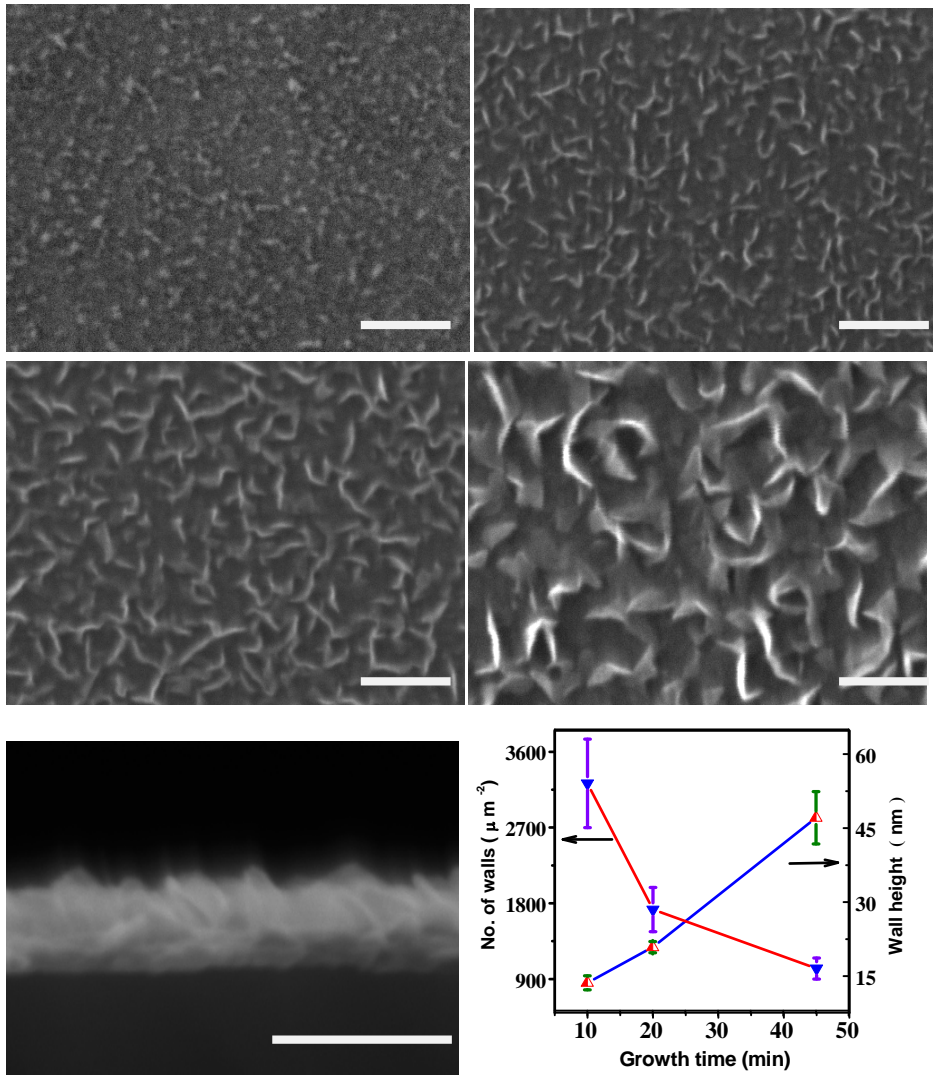

Fig. 2- The SEM micrographs of the VGNs grown at different duration of (a) 10 sec (b) 10 mins (c) 20 mins and (d) 45 mins. (e) cross-sectional SEM micrograph of 45 mins grown sample (f) The plot indicates the wall density and wall height as a function of growth duration. The scale bar in images (a-e) is 100 nm.

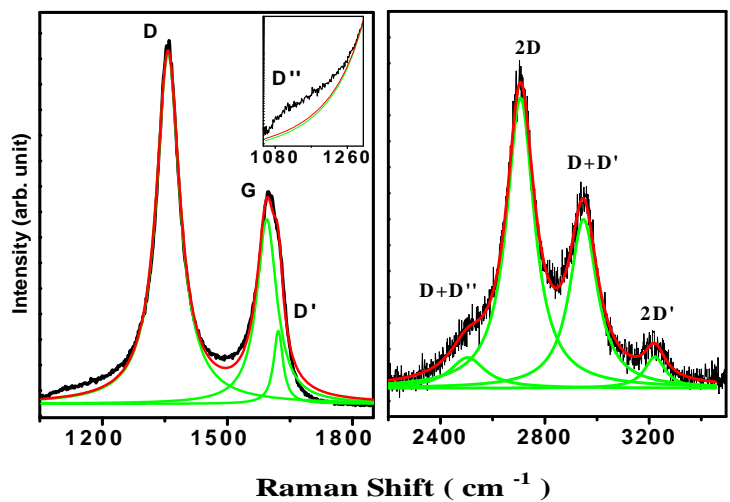

Fig.3- Typical Raman spectrum of VGNs grown on $SiO_2$/Si by ECR-CVD and its deconvolution with Lorentzian line shape.

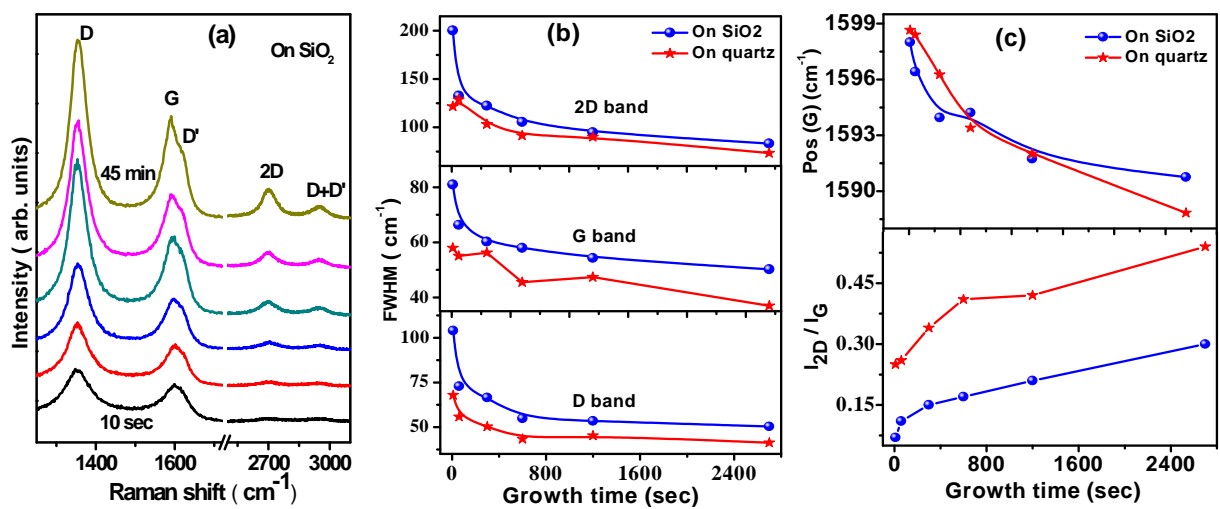

Fig.4- (a) Evolution Raman spectra of VGNs at different growth duration of 10 sec, 1, 5, 10, 20 and 45 mins. The spectra are stacked up vertically with increasing growth time (b) FWHM of D, G and 2D- bands and (c) The variation of G-band position and $I_{2D}/I_G$ with respect to growth time. The solid lines in Fig4b, 4c are guide to eye.

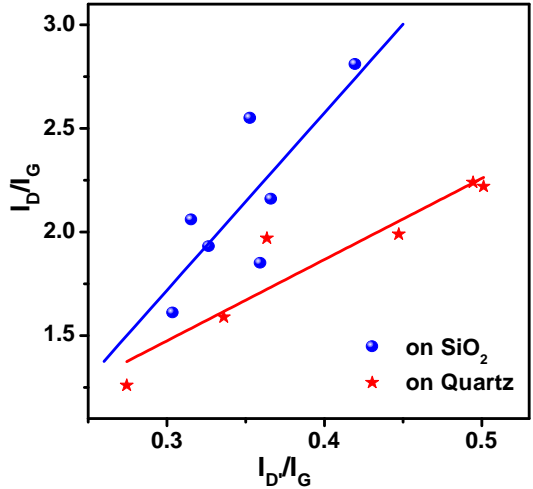

Fig.5- The plot of $I_D/I_G$ vs $I_{D'}/I_G$ for the VGNs grown on $SiO_2/Si$ and quartz substrates. The slope, 8.57±2.86, indicates combined vacancy-type and hopping defects for films grown on $SiO_2/Si$, and 3.98±0.68 represents boundary-like defects for the film on quartz substrates.

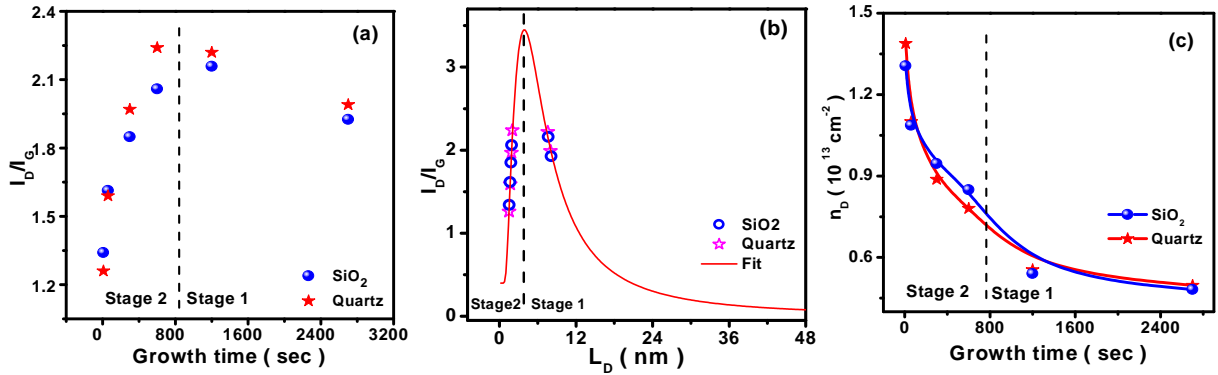

Fig.6- (a) The variation of $I_D/I_G$ with respect to growth time (b) the plot of $L_D$ versus $I_D/I_G$ (c) the calculated defect density with respect to growth time for the VGNs grown on quartz and $SiO_2/Si$ substrates. The solid line in the Fig. 6b is the fit to the experimental data along with simulated curve with the same parameters for $L_D$ ranges from 0 to 50 nm. The fitting is performed using equation 5. The solid line in the Fig. 6c is guide to eye.

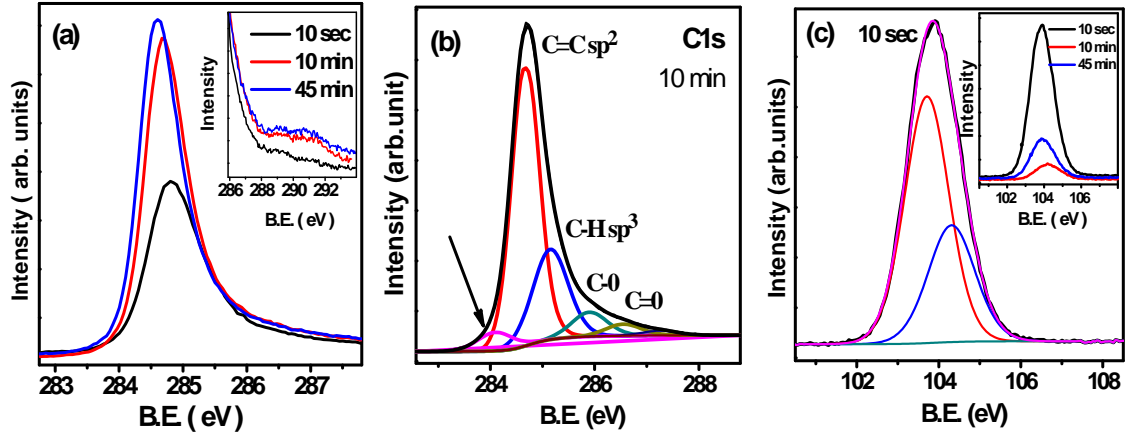

Fig.7- (a) the high resolution C1s spectra for the films grown for 10 sec, 10 min and 45 min (b) deconvulated C1s spectrum for the film grown for 10 min and (c) the Si2p XPS spectrum for 10 sec grown film. The inset in Fig. 7a is an extended region of C1s at higher B.E. The inset in Fig. 7c is the combined Si2p spectra for the three films studied here.

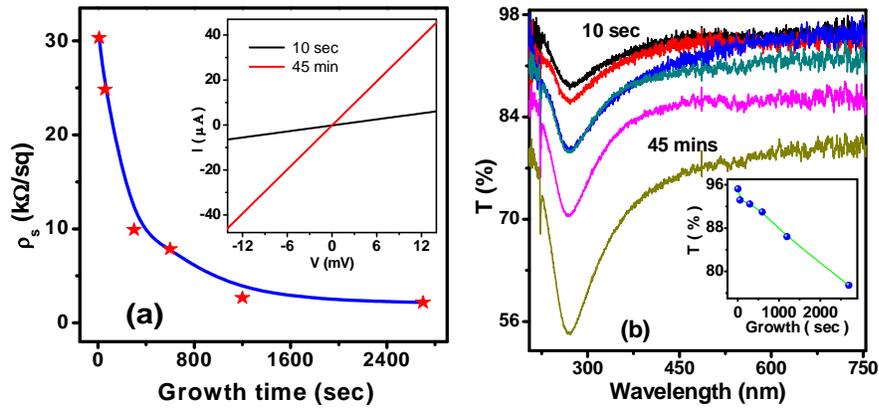

Fig.8. (a) The variation of sheet resistance of VGNs grown on $SiO_2$/Si substrate and (b) UV-Vis transmission spectra of VGNs grown on quartz substrate with respect to growth time. The inset in Fig. 8(b) shows optical transmittance at 550 nm. The solid line in Fig. 8a is guide to eye.

19